\documentclass[fleqn,usenatbib]{mnras}

\usepackage[T1]{fontenc}
\usepackage{ae,aecompl}

\usepackage{graphicx}	
\usepackage{amsmath}	
\usepackage{amssymb}	
\usepackage{multicol,multirow}
\usepackage{booktabs,tabularx}

\newcommand{\hst}{\textit{HST}} 
\newcommand{\spitzer}{\textit{SST}} 
\newcommand{\msun}{\text{M}_\odot} 
\newcommand{\lsun}{\text{L}_\odot} 
\newcommand{\rsun}{\text{R}_\odot}

\def\gax{\mathrel{\raise.3ex\hbox{$>$}\mkern-14mu\lower0.6ex\hbox{$\sim$}}}
\def\lax{\mathrel{\raise.3ex\hbox{$<$}\mkern-14mu\lower0.6ex\hbox{$\sim$}}}
\def\gtorder{\mathrel{\raise.3ex\hbox{$>$}\mkern-14mu
             \lower0.6ex\hbox{$\sim$}}}
\def\ltorder{\mathrel{\raise.3ex\hbox{$<$}\mkern-14mu
             \lower0.6ex\hbox{$\sim$}}}

\title[No pre-SN outbursts from SN\,2023ixf]{Constraints on pre-SN outbursts from the progenitor of SN\,2023ixf using the Large Binocular Telescope}

\author[Neustadt et al.]{
J.~M.~M.~Neustadt,$^{1}$\thanks{E-mail: neustadt.7@osu.edu (JMMN)}
C.~S.~Kochanek,$^{1,2}$ 
M.~Rizzo~Smith$^{1}$
\\
$^{1}$Department of Astronomy, The Ohio State University, 140 West 18th Avenue, Columbus, OH 43210, USA \\
$^{2}$Center for Cosmology and AstroParticle Physics (CCAPP), The Ohio State University, 191 W. Woodruff Avenue, Columbus, OH 43210, USA \\ 
}

\date{Accepted XXX. Received YYY; in original form ZZZ}

\pubyear{2023}

\begin{document}
\label{firstpage}
\pagerange{\pageref{firstpage}--\pageref{lastpage}}
\maketitle

\begin{abstract}
The progenitor of SN\,2023ixf was a $\sim$10$^{4.8}$ to $10^{5.0}\,\lsun$ star ($\sim$9 to $14\,\msun$ at birth) obscured by a dusty $\dot{M} \simeq 10^{-5}\,\msun \rm \,yr^{-1}$ wind with a visual optical depth of $\tau_V \simeq 13$.  This is required by the progenitor SED, the post-SN X-ray and H$\alpha$ luminosities, and the X-ray column density estimates.  In Large Binocular Telescope (LBT) data spanning 5600 to 400~d before the SN, there is no evidence for optical variability at the level of $\sim$10$^3\,\lsun$ in $R$ band, roughly 3 times the predicted luminosity of the obscured progenitor.  This constrains direct observation of any pre-SN optical outbursts where there are LBT observations.  However, models of the effects of any pre-SN outburst on the dusty wind show that an outburst of essentially any duration exceeding $\sim$5 times the luminosity of the progenitor would have detectable effects on the dust optical depth for decades. While the dust obscuration here is high, all red supergiants have dusty winds, and the destruction (or formation) of dust by even short-lived transients will always have long term effects on the observed brightness of the star because changes in the dust optical depths after a luminous transient occur very slowly.
\end{abstract}


\section{Introduction}\label{intro}

The life of a massive star (>8\,$\msun$) ends with the collapse of its core, which is sometimes followed by an explosive ejection of its envelope and a luminous transient known as a core-collapse supernova (SN).  An open question is whether the progenitor stars of SNe experience outbursts prior to explosion -- does the star ``signal'' that the core is about to collapse?  

Evidence for pre-SN variability includes the direct observation of outbursts for a number of core-collapse SN progenitors (e.g., \citealt{pastorello07,fraser13,mauerhan13,ofek14,ofek16,jacobson23}) as well as inferences from the class of Type~IIn SNe, which are characterized by narrow emission lines in their spectra.  These narrow lines are indicative of a dense, slow-moving circumstellar medium (CSM) interacting with the SN shock \citep{smith14} that may have originated as ejected material from the red supergiant (RSG) progenitors in the years/decades prior to explosion.  Indeed, pre-SN outbursts are commonly observed in Type~IIn progenitors \citep{ofek14}.  

It may be the case that ``normal'' Type~II-P progenitors also undergo similar but smaller outbursts prior to SN that can explain the properties of the SN lightcurves \citep{morozova17,morozova20,forster18,davies22}.  The physical mechanism(s) producing these outbursts are unclear, as the inferred mass-loss rates are much higher ($\dot{M} \gtorder 10^{-4} \rm \,\msun\,yr^{-1}$, e.g. \citealt{Moriya2014}) than what is achievable via radiation-driven winds typical of RSGs ($\dot{M} \ltorder 10^{-5} \rm \,\msun\,yr^{-1}$, e.g. \citealt{beasor20}), but there are many possibilities (e.g., \citealt{ouchi19,matsumoto22,ko22,Tsuna2023}), including luminosity-driven gravity waves originating in the core that heat the outer envelope \citep{quataert12,Fuller2017,wu21}.  There are also arguments that the mass-loss rates of these outbursts are significantly overestimated because they do not account for the wind acceleration \citep{moriya17} or the complex atmospheres of RSGs \citep{dessart17,goldberg22}.

If such outbursts are common, one could survey massive stars for signs of pre-SN variability and predict their imminent death.  One such survey is the search for failed SNe with the Large Binocular Telescope (LBT, \citealt{hill06}), first proposed by \citet{kochanek08}.  This survey monitors luminous stars in 27 galaxies within 10~Mpc using the LBT and is designed to detect the death of evolved $\sim$9--30~$\msun$~stars independent of whether they explode as SNe. Papers discussing this survey for failed SNe and its candidates include \citet{gerke15}, \citet{adams17a}, \citet{adams17b}, \citet{basinger20}, and \citet{neustadt21}.  

This survey can also be used to study the variability of stars which do explode \citep{szczygiel12,johnson17,kochanek17,johnson18}.  In particular, \citet{johnson18} examined the progenitor lightcurves of four Type~II SNe and found that all of them were quiescent at the $\sim$10~per~cent level in the 5--10~yr prior to the SN.  Furthermore, \citet{johnson18} used this result to estimate that no more than 37~per~cent of normal Type~II (i.e., not Type~IIn) progenitors could have an extended outburst in the years prior to the SN.  \citet{rizzo23} analyzed the late-time evolution of twelve core-collapse SNe and searched for evidence of extreme mass-loss episodes in the decades prior to explosion -- these would be detectable in the late-time lightcurves as the SN shock encountered the expanding mass-loss shell.  No such extreme outbursts were detected; instead, the luminosities were consistent with ``normal'' RSG mass-loss rates of $-7.9 \leq \log \rm (\msun\,yr^{-1}) \leq -4.8$.   

SN\,2023ixf is a Type~II core-collapse SN \citep{perley23} discovered by K.~Itagaki on 2023 May 19 (MJD\,60083.7) in the nearby spiral galaxy M101 (NGC~5474, $\alpha = \rm 14^h 03^m 38\fs51$, $\delta =+54\degr18\arcmin42\farcs10$).  \cite{szalai23} identified a dusty progenitor in archival \textit{Spitzer Space Telescope} (\spitzer) observations and found no evidence for mid-infrared (IR) variability.  The progenitor has also been identified in archival \textit{Hubble Space Telescope} (\hst) observations \citep{Soraisam2023,Pledger2023} at F814W but not at F555W or F475W.  The SN also has heavily absorbed X-ray emission (\citealt{Grefenstette2023}, \citealt{Kong2023}, \citealt{Mereminskiy2023}, \citealt{chandra23}) and transient, narrow H$\alpha$ emission \citep{Yamanaka2023}, both of which are indicative of a fairly dense CSM.  We adopt a distance to M101 of 6.14~Mpc \citep{shappee11}.

Here we report the pre-SN LBT lightcurve of SN\,2023ixf.  In Section~\ref{sec:progenitor}, we discuss the properties of the progenitor, finding that the progenitor was obscured by a fairly dense wind that is sufficient to explain the X-ray and H$\alpha$ observations of the SN.  In Section~\ref{sec:im_sub}, we discuss our image subtraction methods and present the progenitor LBT lightcurve.  In Section~\ref{sec:dust}, we discuss how pre-SN outbursts can be constrained through their effects on the dust obscuration. In Section~\ref{sec:conclusion}, we summarize our findings and comment on possible mass-loss episodes missed by our observations.  

\section{The Progenitor Star And Its Mass Loss}\label{sec:progenitor}

Here we discuss the properties of the progenitor star of SN\,2023ixf based on both archival IR and optical observations and the X-ray and H$\alpha$ emission of the SN.  \cite{szalai23} reports \spitzer\ 3.5 and 4.5$\,\mu$m progenitor fluxes of $17.6\pm0.2$ and $17.2\pm0.2$~mag with no detections at the significantly less sensitive 5.8 and 8.0$\,\mu$m bands.  \citet{Pledger2023} report a \hst\ F814W detection of $24.41\pm0.06$~mag (\citealt{Soraisam2023} reports $24.39\pm0.08$) and detection limits of $26.7$ and $26.6$~mag for F435W and F555W.  We model this spectral energy distribution (SED) following \cite{Adams15} using {\tt DUSTY} \citep{Ivezic1997,Ivezic1999,Elitzur2001} using Solar metallicity MARCS \citep{Marcs2008} at lower temperatures (2600 to 4000\,K) and \citet{Castelli2003} stellar atmosphere models and higher temperatures, with \cite{Draine1984} silicate circumstellar dust.  Given the limited constraints, we used a temperature prior for RSGs of $T_* = 3800\pm 350$\,K based on \cite{Levesque2005}. As noted by \cite{szalai23}, the high \spitzer\ fluxes are indicative of dust emission, while the blue colors of the SN suggest little host extinction (the Galactic extinction is negligible, $A_V=0.024$, \citealt{Schlafly2011}), so we consider models with no foreground extinction and fixed dust temperatures of $T_d=1500$, 1250, 1000, and $750$\,K. 

With only three fluxes, we obtain essentially perfect fits for all four dust temperatures and cannot distinguish between them.  The limits on the stellar temperature simply replicate the prior, the stellar luminosities of the models span the range $10^{4.75}$ to $10^{5.00}\,\lsun$, and the visual optical depth is $\tau_V=12.5$ with uncertainties that rise from $\pm 0.7$ for the hottest dust temperatures to $\pm 1.5$ for the coldest. The inner edge of the dust distribution simply depends on the choice of the dust temperature with $R_d =22\pm3$, $33\pm4$, $52\pm 8$, and $85\pm 18~\rm AU$ going from the hottest to coldest dust temperatures.  For the Solar metallicity PARSEC isochrones (\citealt{Bressan2012}, \citealt{Marigo2013})  the final luminosity is roughly related to the initial mass by $\log L_* = 4.8 + 1.5\log(M_*/10\,\msun)$, so this luminosity range corresponds to an initial mass range of $9.3$ to $13.6\,\msun$.  The final luminosities of the \citet{Groh2013} progenitor models give similar results.  If we include no dust and simply normalize the progenitor to match the F814W luminosity at the same stellar temperature, the luminosity ($10^{4.26}\,\lsun$) corresponds to a star with too small a mass ($\sim$4$\,\msun$ for the scaling above) to explode as an SN.

\begin{figure}
\includegraphics[width=0.95\linewidth]{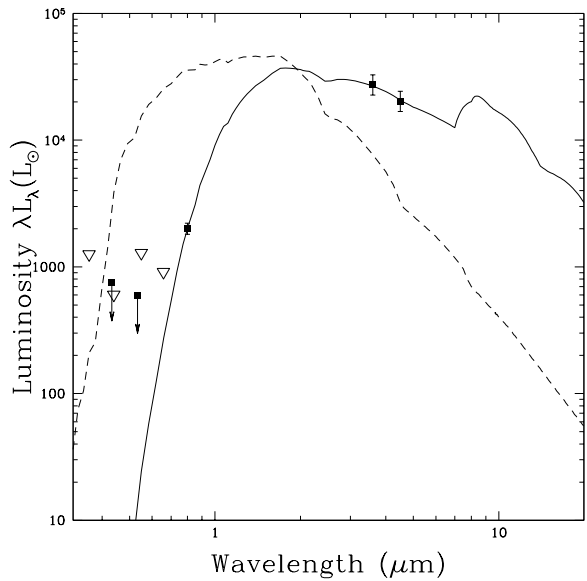}
\caption{ The observed SED (solid line) of the progenitor as fit to the measured luminosities (filled squares) from \citet{Pledger2023}, \citet{szalai23} and \citet{Soraisam2023}.  The dashed line shows the unobscured SED of the progenitor for comparison.  The open, downward pointing triangles are the RMS variability limits found from the LBT observations for the \textit{UBVR} bands (See Sec.~\ref{sec:im_sub}).}
\label{fig:sed}
\end{figure}

We used a $R_{\rm out}/R_{\rm in}=2$ dusty shell with a $\rho \propto 1/r^2$ density profile, but extending the radial range or changing to graphitic dusts will have little effect given the limited constraints.  For scaling our subsequent models, we adopt (in round numbers) $T_*=3900 T_{*0}$\,K, $L_* = 10^{4.8} L_{*0}\,\lsun$, $R_* = 550 R_{*0}\,\rsun = 550 L_{*0}^{1/2} T_{*0}^{-2}\,\rsun$, $T_d=1250 T_{d0}$\,K, $\tau_V=13 \tau_0$ and $R_{\rm in}= 7000 R_{\rm in0}\,\rsun$ since this is roughly the condensation temperature of silicate dusts (see the review of \citealt{Tielens2022}).  The subscript $0$ quantities track the dependence of the results on the parameters (so $L_{*0}=L_*/10^{4.8}\,\lsun$). Figure~\ref{fig:sed} shows the $T_d=1250$\,K fits to the observed SED as well as the unobscured SED of the progenitor.

Given these parameters, we can then estimate the physical properties of the required wind and its effect on post-SN observables, including the X-ray luminosity, column density, and H$\alpha$ luminosity.  We scale the wind velocity to $v_w = 10 v_{w0}\rm \,km\,s^{-1}$, the SN shock velocity to $v_s=5000 v_{s0}\rm \,km\,s^{-1}$ and the visual dust opacity of the wind to $\kappa_V = 100 \kappa_0$\,cm$^2$\,g$^{-1}$. For a $\rho\propto r^{-2}$ wind, these parameters imply a wind mass-loss rate of 
\begin{equation}
\dot{M}= { 4 \pi v_w R_{d} \tau_V \over \kappa_V }
   = { 1.3 v_{w0} R_{d0} \tau_0 \over \kappa_0} \times 10^{-5}\,\msun \rm \,yr^{-1} \,.
\end{equation}
This is on the high end for RSGs (e.g., \citealt{beasor20}), but nowhere near the mass loss rates invoked for luminous Type~IIn SNe ($\dot{M}>10^{-4}\,\msun$\,yr$^{-1}$, e.g., \citealt{Moriya2014}).   As is usual for winds, the quantity that is actually fixed is $\dot{M}/v_w$ and not either quantity independently. Note that $R_{\rm in}/v_w=15 R_{\rm in0}/v_{w0}$\,yr, so the dusty wind has to have been in existence for longer than the period spanned by the LBT observations.  The intense shock break-out luminosity spike (e.g., \citealt{Ensman1992}) of the SN would evaporate all the dust at these distances, so one would expect essentially no dust absorption of the optical/UV SN emission from this material.

A SN shock moving through such a wind generates a luminosity of
\begin{equation}
       L_s = { \dot{M} v_s^3 \over 2 v_w }
           =  (5.0 \times 10^{40}) R_{\rm in0} \tau_0 v_{s0}^3 \kappa_0^{-1} \rm \,erg \, s^{-1},
\end{equation}
so the observed X-ray luminosities -- $1.1\times 10^{40}$\,erg\,s$^{-1}$ in \cite{Grefenstette2023}, $3.8\times 10^{39}$\,erg\,s$^{-1}$ in \cite{Kong2023}, $1.7 \times 10^{40}$\,erg\,s$^{-1}$ in \cite{Mereminskiy2023} and $8 \times 10^{39}$\,erg\,s$^{-1}$ in \cite{chandra23} -- can be supported if 10--20~per~cent of the shock luminosity is being radiated as X-rays.  If the shock front is located at radius $R_s = R_*+v_s t$, then the wind column density outside the shock is $N_{\rm H}=4.7$, $3.1$, $2.3$, $1.8$ and $1.5 \times 10^{23}$\,cm$^{-2}$ on days 1 through 5 assuming a pure hydrogen wind, which is exactly the scale inferred from the X-ray absorbing column density of $2 \times 10^{23}$\,cm$^{-2}$ on day~3 (MJD~60086.7, \citealt{Grefenstette2023}).  By day~11, it would only be $7 \times 10^{22}$\,cm$^{-2}$. This is qualitatively consistent with the drop to $3\times 10^{22}$\,cm$^{-2}$ seen by \cite{chandra23} on day~11.9 (MJD~60095.6), although better quantitative agreement would be found using a higher shock speed.

In addition to the SN shock break out radiation pulse destroying the dust, it will also photoionize the wind, leading to narrow recombination lines in the spectra (``flash spectroscopy'', e.g., \citealt{Khazov2016,Yaron2017,Kochanek2019}).  The luminosity scale of the H$\alpha$ emission is
\begin{equation}
L_{\rm H\alpha}= {\dot{M}^2 \alpha_{\rm H\alpha} E_{\rm H\alpha} \over 4 \pi v_w^2 m_p^2 r_s  } \,, 
\end{equation}
again assuming a pure hydrogen wind, where $\alpha_{\rm H\alpha} E_{\rm H\alpha}=3.54 \times 10^{25}$\,erg\,cm$^3$\,s$^{-1}$ at $10^4$\,K \citep{Draine2011}.  The actual time dependence is more complex because of light travel times and the expansion of the ionizing radiation pulse (see \citealt{Kochanek2019}), but this is an adequate estimate given the available information.  Note that broad wings on the narrow lines are produced by radiative acceleration of the wind and do not require a Thomson optically thick medium to create them through scattering \citep{Kochanek2019}. This implies H$\alpha$ luminosities of $7.9$, $5.2$, $3.9$, $3.1$ and $2.5 \times 10^{37}$\,erg\,s$^{-1}$ on days 1 through 5.  We fit the day~1.2 (MJD~60084.9) spectrum from \citet{teja23} and found a narrow H$\alpha$ line flux of $2.3 \times 10^{-13}$\,erg\,s$^{-1}$\,cm$^{-2}$ or $9 \times 10^{38}$\,erg\,s$^{-1}$, so these estimates are in the right regime given how sensitive the line flux is to $\dot{M}/v_w$ and the radius of the RSG.

In summary, the progenitor of SN~2023ixf was a relatively low mass RSG heavily obscured by a relatively dense, but not extraordinary, dusty wind.

\section{The Progenitor lightcurves}\label{sec:im_sub}

\begin{table*}
\caption{RMS Luminosity and Slopes}\label{tab:lc}
\begin{tabularx}{0.66\linewidth}{p{0.03\textwidth}>{\raggedleft\arraybackslash}X>{\centering}p{0.08\textwidth}>{\centering\arraybackslash}p{0.10\textwidth}>{\centering\arraybackslash}p{0.12\textwidth}>{\centering\arraybackslash}p{0.12\textwidth}} \toprule
\multirow{2}{*}{Band} & \multirow{2}{*}{Epochs} & \multicolumn{2}{c}{RMS Luminosity [$10^3\,\lsun$]} & \multicolumn{2}{c}{Slope [$10^3\,\lsun \rm \,yr^{-1}$]} \\
 & & $\sigma_{\rm SN}$ & $\langle \sigma_{i} \rangle$ & $\hphantom{-} \beta_{\rm SN}$ & $ \hphantom{-}\langle \beta_{i} \rangle$  \\  \midrule \midrule
$\hphantom{,}R$ & 30 & 0.94 & $0.43\pm0.11$ & $-0.01\pm0.04$ & $-0.02\pm0.03$ \\
$\hphantom{,}V$ & 8 & 1.36 & $1.19\pm0.36$ & $\hphantom{-}0.06\pm0.10$ & $\hphantom{-}0.03\pm0.08$ \\
$\hphantom{,}B$ & 10 & 0.65 & $0.70\pm0.19$ & $\hphantom{-}0.06\pm0.06$ & $\hphantom{-}0.04\pm0.07$ \\
$\hphantom{,}U$ & 8 & 2.45 & $2.36\pm1.14$ & $\hphantom{-}0.23\pm0.07$ & $\hphantom{-}0.11\pm0.13$  \\ \bottomrule
\end{tabularx}
\begin{flushleft}
\textit{Notes:} Band, number of epochs, RMS of SN ($\sigma_{\rm SN}$), mean RMS of background points ($\langle \sigma_{i} \rangle$), slope of SN ($\beta_{\rm SN}$), and mean slope of background points ($\langle \beta_{i} \rangle$). The error bounds of the mean RMS and mean slope refer to the RMS of the associated quantities.  The error bounds of the slope of SN refer to the standard error of the slope as computed from a least squares fit.
\end{flushleft}
\end{table*}

Here we discuss the 15~yr of optical lightcurve of the progenitor of SN\,2023ixf from the LBT search for failed SN.  The Large Binocular Camera (LBC, \citealt{giallongo08}) on the LBT consists of four 2048$\times$4608~pixel (7.8$\times$17.6~arcmin) detectors -- a central chip with two chips adjacent and parallel to the long axis and one chip perpendicular to the short axis with 18~arcsec gaps between them.  The survey makes use of LBT's unique binocular feature, where we observe in the $R$ filter with the red optimized LBC-Red camera while simultaneously cycling through observations in the $V$, $B$ and $U_{\rm spec}$ (hereafter $U$) filters with the blue optimized LBC-Blue camera.  Unfortunately, SN\,2023ixf sits near the edge of the central chip, so some of our epochs have the SN location in the gap.  For $R$ band, this was only 5 images out of the total 44, but for $V$, $B$, and $U$ bands, we had to exclude roughly 30 images out an average of 40.  Unfortunately, no data were obtained in Spring 2023 due to poor weather conditions at Mt.~Graham\footnote{We would have likely observed M101 in the 2023 May 15--18 run had the weather cooperated.}, and the last usable epoch is from 2022 February 10 (MJD~59620, 463~d prior to SN).

We use the \texttt{ISIS} image subtraction package \citep{alard98,alard00} with the same astrometric references as those used in \citet{gerke15}.  We ran image subtraction on all epochs to construct lightcurves, and we flag epochs with FWHM~>~1.5~arcsec to exclude epochs with bad seeing and image subtraction scaling factors <~0.8 to exclude observations with significant cirrus.  The number of unflagged epochs is given in Table \ref{tab:lc}.  The image subtraction process requires constructing a ``reference'' image in each filter that we scale and subtract from the individual epochs.  The reference image is constructed using observations with the best seeing conditions (< 1~arcsec).  For $R$ band, we are able to use the most recently updated reference image from \citet{neustadt21}.  For the $V$, $B$, and $U$ bands, we made new reference images constructed only from images that contain the SN position, leading to reference images that were noisier than in \citet{neustadt21}.  

\begin{figure}
\includegraphics[width=0.95\linewidth]{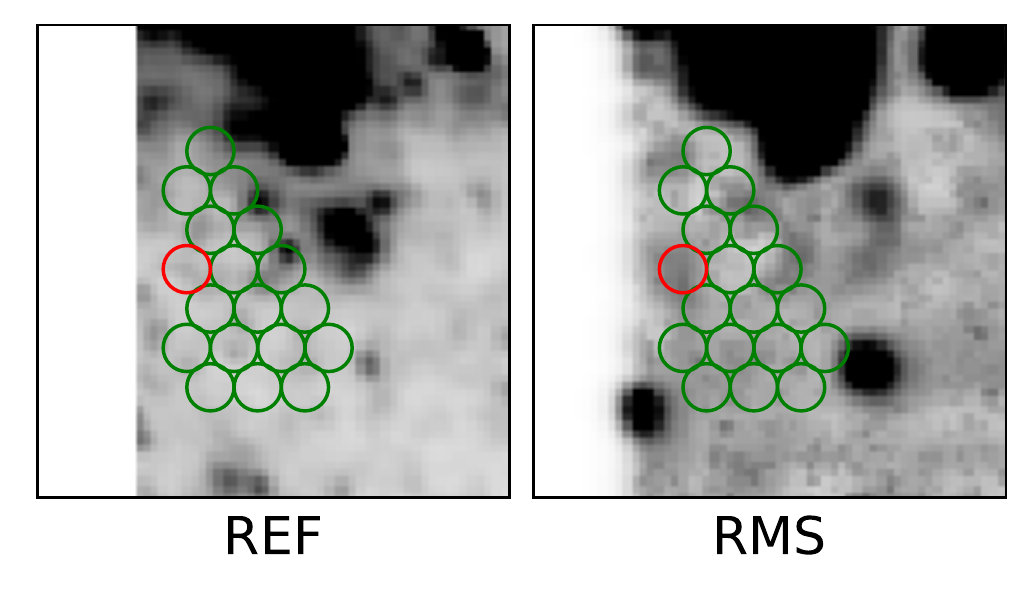}
\caption{$R$-band imaging of the progenitor showing the SN location and background regions as red and green circles, respectively.  Each circle is 3 pixels (0.7~arcsec) in radius separated by 5.8 pixels (1.3~arcsec).  The white space to the left of the SN location is due to the variable location of the field with respect to the chip gap.  The reference (REF) and RMS images do not show an obvious point source at the SN location.  The RMS image combines the \texttt{ISIS} subtracted images such that each pixel in the RMS image is the RMS of that position's pixel values in all the subtracted images.  This process highlights variable sources.  Note that this is different from the RMS values discussed in Sec.~\ref{sec:im_sub}.}
\label{fig:bck}
\end{figure}

As in \cite{johnson18}, we extract lightcurves both for the SN progenitor and for a grid of nearby points to measure background fluctuations, shown in Figure~\ref{fig:bck}.  We use these background lightcurves to empirically estimate the errors in the progenitor lightcurves, since the errors reported by \texttt{ISIS} tend to be underestimates.  We calibrate the data as in \citet{gerke15} and \citet{adams17b}.  Sloan Digital Sky Survey (SDSS, \citealt{ahn12}) stars with SDSS \textit{ugriz} AB magnitudes are matched with stars in the reference images and transformed to \textit{UBVR} Vega magnitudes using the conversions reported by \citet{jordi06} and zero-points reported by \citet{blanton07}.

\begin{figure*}
\includegraphics[width=\linewidth]{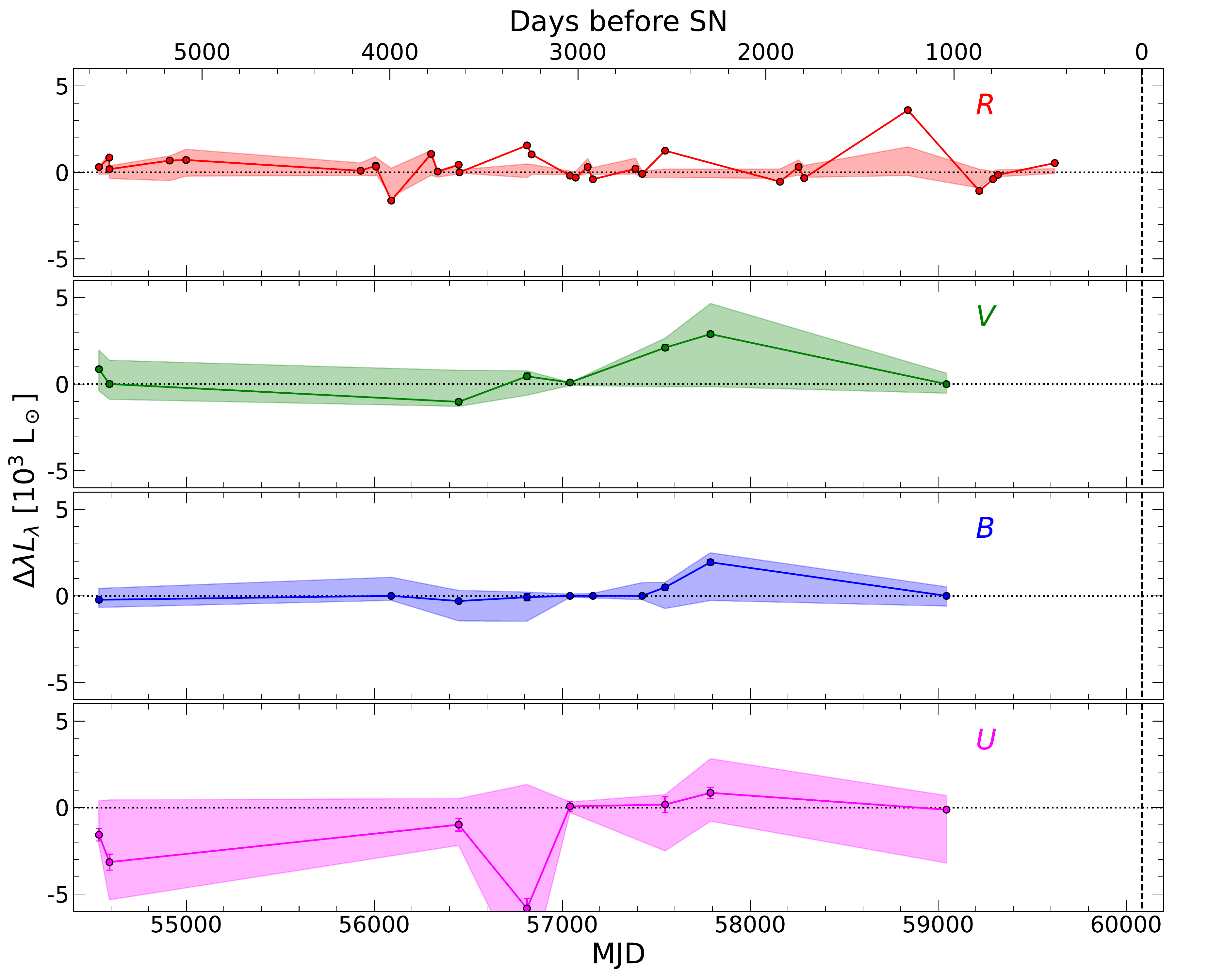}
\caption{Difference image \textit{UBVR} lightcurves of the progenitor of SN\,2023ixf.  Differential luminosity ($\Delta \lambda L_\lambda$) is measured relative to the reference image.  The shaded region is the 1$\sigma$ scatter about the mean of the 17 nearby background comparison points.}
\label{fig:lc}
\end{figure*}

In Figure~\ref{fig:lc}, we present the subtracted \textit{UBVR} lightcurve of the progenitor as band luminosities ($\lambda L_\lambda$).  These are not corrected for the negligible Galactic extinction
or for host extinction.  In $R$ band, most of the observations do not exceed the RMS of the background fluctuations.  There are a few that do (e.g., the three observations on MJD~56812, 57547, and 58838), but when we look at the subtracted images for these epochs (not shown), these are very likely to be subtraction artifacts.

In Table~\ref{tab:lc}, we report the root-mean-square (RMS, $\sigma$) value of the lightcurves of the progenitor and the mean RMS value of the array of background points.  We also include these RMS values as upper limits in the flux in each band in Figure~\ref{fig:sed}.  In the $U$ and $B$ bands, the SN RMS does not exceed that of the background, and in the $V$ band it only marginally exceeds it.  We do see that in the $R$ band, the RMS of the progenitor is roughly 2 times the mean background RMS, but this is driven by outliers. For example, if we remove the two points at MJD~57547 and 58838, the progenitor and mean background RMS drops to 0.66 and $0.42\times10^3\,\lsun$, respectively. 

We also made linear fits to the lightcurves of the progenitor and the array of background points.  The slopes ($\beta$) of the progenitor lightcurves, their associated errors, the mean slopes of the background points, and their associated RMS's are presented in Table~\ref{tab:lc}.  For each quantity, the measurement errors from \texttt{ISIS} are not included in the calculations.  In the $R$, $V$, and $B$ bands, the slopes of the progenitor lightcurve are consistent with zero.  In the $U$ band, the slope is nonzero but is comparable to the mean slope of the background points.  Combining the measurements of the slopes of the progenitor lightcurves along with the RMS of the lightcurves, we have strong evidence that the progenitor of SN\,2023ixf did not vary coherently or stochastically in the 5600~d (15~yr) prior to explosion. 

As we can see in Figure~\ref{fig:sed}, the RMS limits on the flux variability are much larger than the predicted flux from the obscured SED.  For example, the $R$-band RMS limit is $\sim$3 times larger than the predicted $R$-band luminosity.  Unlike the earlier LBT results \citep{johnson18}, we are only sensitive to modestly luminous outbursts and far from being able to detect the normal variability of RSGs.  There is the question of whether a transient can be fit into the gaps in the LBT lightcurves, but it is difficult for stars the size of RSGs to have short transients other than shock break out pulses.  For example, the pre-SN outburst models of \citet{Fuller2017} or \citet{Tsuna2023} have extended ($\sim$year) luminous transients after a shorter, initial luminosity impulse.  As we next discuss, the presence of a dusty CSM means that any luminous transient produces very long lasting effects ($\sim$decades) on the observed brightness of the progenitor through the effects of the transient on the dust optical depth.

\section{The Effect of a Radiation Pulse on the Dusty Wind} \label{sec:dust}

If the star undergoes a luminous transient before the SN, there is the obvious effect of the source becoming brighter.  However, with a dusty wind, the bigger effect on the observed optical luminosity can be from the change in the optical depth of the wind, since changes in the optical depth exponentially affect the escaping radiation.  Changes in the dust also mean that short-duration transients have long-lived consequences for the observed optical luminosity.  For example, if it was simply a matter of waiting to move the dusty material to a larger radius, the characteristic time scale for material at radius $R_d$ would be $R_d/v_w$, which is five years for $R_d=10$\,AU and $v_w=10$\,km/s. In practice, dust can reform in place, and it is easiest to illustrate those time scales with simulations. 

The temperature $T_d$ of a dust grain of radius $a$ is determined by the radiative balance between absorbed and radiated energy,
\begin{equation} \label{eqn:teq}
{ L_* \over 4 \pi r^2} \pi a^2 Q_{\rm abs}(T_*) = 4\pi a^2 Q_{\rm em}(T_d) \sigma T_d^4, 
\end{equation}
where $r$ is distance from the star and $Q_{\rm abs}(T_*)$ and $Q_{\rm em}(T_d)$ are the Planck averaged absorption and emission factors (e.g., \citealt{Waxman2000}).  We are assuming moderate optical depths where the IR emission of the grains simply escapes. The ratio of these factors for \cite{Draine1984} (astro)silicate dust grains smaller than $0.3\mu$m and a source temperature of $T_*=4000$\,K is well modelled by
\begin{equation}
 {Q_{\rm em}(T_d) \over Q_{\rm abs}(T_*) } \simeq { c_0 \hat{T}_d^2 \over 1 + c_1 \hat{T}_d^3 }
       { 1 \over 1 + c_2 a }
\end{equation}
where $\hat{T}_d=T_d/1000$\,K, $c_0=39.0$, $c_1=59.1$ and $c_2=6.27\mu$m$^{-1}$.  This makes the energy balance equation a quadratic in $T_d^3$ with only one physical solution.  As the first of our simplifications, we will neglect the contribution to the energy balance from the evaporation and condensation onto grains -- based on \cite{Waxman2000}, these appear to only be important when the grain is so close to evaporation that it is a minor correction.  

We consider dust formed in a $\rho\propto 1/r^2$ wind where the growth rate can be modelled as 
\begin{equation}
{ d a \over dt } = { v_w a_\infty \over r^2 } - 
\nu_0 \left( { m_d \over \rho_d} \right)^{1/3}\exp\left(-Q/k T_d \right)
\end{equation}
where $m_d$ is the mass of a dust monomer ($140 m_p$ for Mg$_2$SiO$_4$), $\rho_d$ is the bulk dust density ($\rho_d=3$\,g\,cm$^{-3}$), and $\nu_0 = 2\times 10^{15} \rm \,s^{-1}$ and $Q/k=68100$\,K describes the evaporation rate of the dust \citep{Waxman2000}. The first term is the collisional growth of the grain and the second term is its evaporation rate. The dust formation radius $R_d$ is the point where $da/dt \equiv0$ in Equation~\ref{eqn:teq}. We can largely ignore evaporation for $r>R_d$, in which case the grain size is
\begin{equation}
a = a_\infty \left(1 - { R_d \over r } \right) 
\end{equation} 
with an asymptotic grain radius of
\begin{equation}
a_\infty = { \dot{M} f_s f_g v_c \over 16 \rho_d v_w^2 R_d}
\end{equation}
where $f_s$ is the probability of a collision sticking, $f_g$ is the mass fraction of the wind in the condensible species and $v_c$ is the collision velocity. This somewhat simplifies the collisional growth because it assumes that the collisional rates for merging grains are the same as the rate for a grain colliding with the same number of monomers. This is conservative since it will overestimate the ability of grains to reconstitute themselves after being evaporated by a transient. If we take our standard parameters and use $f_s=1/2$, $f_g=1/200$ and $v_c=v_w/10$, then $a_\infty = 0.08~\mu$m  For our simulations we will use $a_\infty=0.1\,\mu$m as a simple round number.

\begin{figure}
\includegraphics[width=0.95\linewidth]{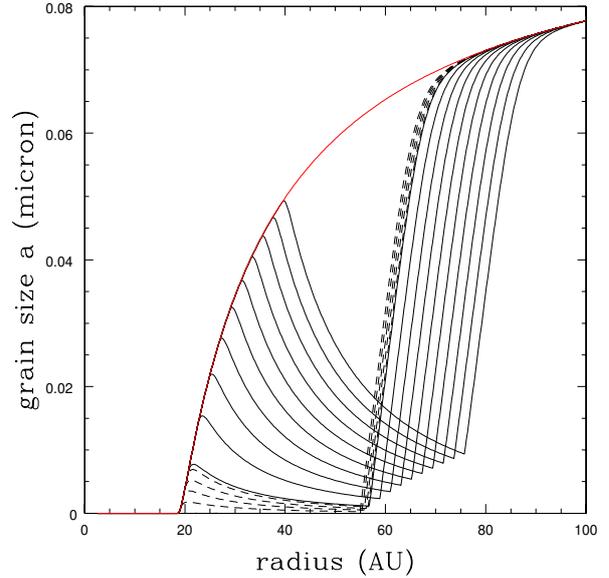}
\caption{Evolution of the grain sizes as a function of radius and time for the $L_{\rm peak}=10L_*$, $\Delta t=0.1$~yr model.  For $L_*$, the dust formation radius is just inside $20$\,AU and the grain sizes for the quiescent wind follow the red ``parabolic'' envelope.  The transient destroys the dust to $\sim$60\,AU and then the dashed lines show the evolution after $0.1$, $0.3$, $0.5$, $0.7$ and $0.9$~yr. The solid lines continue the evolution in yearly intervals out to 10~yr.}
\label{fig:growth}
\end{figure}

Figure~\ref{fig:growth} shows the evolution of the grain sizes for a transient starting at time zero and lasting $\Delta t =0.1$~yr with a peak luminosity of $L_{\rm peak}=10L_*$. The radiation temperature is fixed to $T_*=4000$\,K. For a luminosity of $L_*$, the dust formation radius is a little less than $20$\,AU.  The transient rapidly evaporates the dust to $\sim$60\,AU.   When the transient is over, evaporation is again largely irrelevant outside $\sim$20\,AU, so the subsequent evolution is controlled by the collisional growth rates with two regimes. At large radii, the low densities mean that the grains grow very slowly moving out in radius faster than they grow back to the size they would have at that radius for the original steady state wind.  Collisional growth is fastest at small radii, so the radial size distribution of the steady state wind is steadily restored starting from the inside out.  The net effect is a ``notch'' in the radial size distribution that steadily moves outward and slowly fills in.  

In this analysis we have not included any radiation transport where the dust at smaller radii can shield the grains at larger radii from the transient (or the star).  This is likely safe to do because the evaporation rate varies exponentially with temperature, and the highest temperatures are at the smallest radii. This means that the evaporation times for the dust doing the shielding are shorter than those for the dust being shielded.  So the effect of ignoring the shielding is simply to evaporate the more distant dust a little too fast, which would slightly modify the detailed structure of the grain size distribution in Figure~\ref{fig:growth} near $60$\,AU.

\begin{figure}
\includegraphics[width=0.95\linewidth]{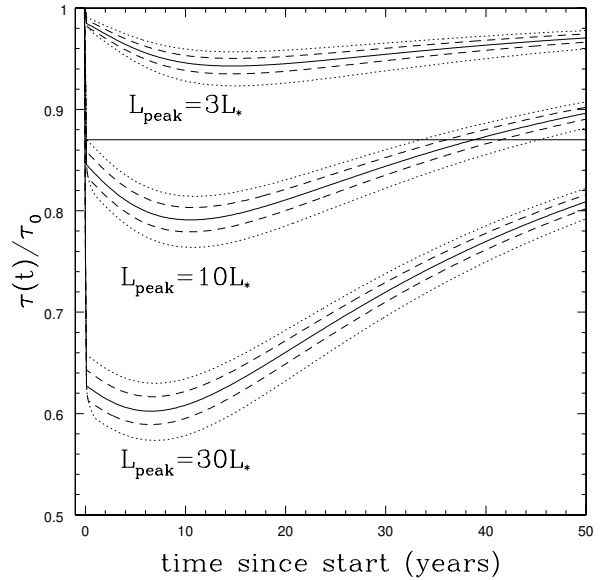}
\caption{The fractional change in the optical depth due to a short radiation spike starting at $t=0$. The peak luminosities are $L_{\rm peak}=3L_*$ (top), $10L*$ (middle) and $30L_*$ (bottom), each shown for durations of $\Delta t =0.01$ (top dotted), $0.03$ (dashed), $0.1$ (solid) $0.3$ (dashed) and $1.0$ (dotted bottom) yr. The horizontal line is approximately the fractional change in the $V$-band opacity needed to produce a detectable signal in the $R$-band observations.}
\label{fig:tau}
\end{figure}

Figure~\ref{fig:tau} shows the evolution of the optical depth for three peak luminosities, $L_{\rm peak}=3L_*$, $10L_*$ and $30L_*$ each modeled as top hats with widths of $\Delta t=0.01$, $0.03$, $0.1$, $0.3$ and $1.0$~years, again fixing $T_*=4000$\,K.  To first approximations, the results are determined by the ratio $L_{\rm peak}/L_*$, and so we only consider $L_* = 10^5\,\lsun$. The initial drop in the optical depth is extremely fast simply because the evaporation rate depends exponentially on the temperature.  This also means that it is the peak luminosity that largely determines the amount of dust destroyed.  The factor of 100 changes in the duration of the transient have only modest effects.   Adding a long luminosity tail to the transient leads to a period where the dust formation radius is larger than for the original wind, which will suppress the growth at the inner edge in Figure~\ref{fig:growth} to larger radii.  Initially, the outward expansion of the surviving dust drives a continued slow decrease in the optical depth which reverses once enough dust has reformed at the inner edge.

Thus, while a short-lived luminous transient can be hidden between the LBT observations, its consequences for the optical depth of the wind are very long lived and so must be observed if large enough.  For a transient to make the progenitor visible in the LBT data through the destruction of dust, we need an increase of the $R$-band luminosity to be above the RMS limit -- roughly a factor of 3 (see Section~\ref{sec:im_sub}).  This would correspond to a decrease in the $R$ band optical depth of $\ln 3 = 1.1$, which roughly corresponds to a drop in the V band optical depth of $1.7$.  This is a fractional change of 13~per~cent for $\tau_V=13$.  For the models in Figure~\ref{fig:tau}, this means that we would have a detectable signal in the LBT data for transients with $L_{\rm peak} \gtorder 5 L_* \sim 5 \times 10^5\,\lsun$ or $2\times 10^{39} \rm \,erg\,s^{-1}$.  This limit is actually much lower than those observed in detected pre-SN outbursts (e.g., \citealt{jacobson23}), and earlier transients cannot be much more luminous because they would still lead to observable optical depth changes over the 15~yr LBT observing period decades after the transient.

\section{Conclusions}\label{sec:conclusion}

Our models for the SED of the progenitor find that it is consistent with a $10^{4.75} \, \lsun \ltorder L_* \ltorder 10^{5.00} \, \lsun$ RSG with a strong $\dot{M}\simeq 10^{-5} \, \msun$\,yr$^{-1}$ wind (for a wind velocity of $v_w=10$\,km\,s$^{-1}$).  This implies a fairly low mass of 9.3 to $13.6 \, \msun$.  The dusty wind was heavily obscuring the progenitor, with a visual optical depth of $\tau_V \simeq 13$.  \cite{szalai23} have already argued for circumstellar obscuration based on the high mid-IR luminosity. Such a dense wind is consistent with the observed X-ray luminosities and column densities as well as the early, narrow H$\alpha$ line emission observed from the SN.

From our 15~yr of LBT observations, we obtain limits on the RMS variability of around $10^3\,\lsun$ and limits on any steady luminosity changes of $<100\,\lsun$\,yr$^{-1}$.  Because of the heavy obscuration, the limits on any pre-SN outburst are weaker than in the previous LBT studies of pre-SN variability \citep{johnson18}, and the limit on the $R$-band RMS variability of $\sim$900$\,\lsun$ is about three times the obscured $R$-band luminosity (see Fig.~\ref{fig:sed}).  

While one can try hiding a short-lived outburst between the LBT epochs, any luminous pre-SN transient leads to a long-lived ($\sim$decades) transient in the wind optical depth.  Since the obscuration is high, and the optical depth exponentially changes the observed fluxes, modest decreases in the optical depth ($\Delta \tau \sim $1--2) would render the star visible in $R$ band.  We illustrate this with simple models with a top hat luminosity transient lasting from $0.01$ to $1.0$~yr, finding that the luminosity is the key variable and the transient duration is of secondary importance.  Roughly speaking, any transient with $L_{\rm peak} \gtorder 5 L_*$ would change the optical depth enough for the previously dust-obscured $R$-band luminosity to become detectable for several years.  Slightly more luminous transients would still lead to detectable optical depth changes over the 15~yr span of the LBT observations even decades after the actual outburst.  While we only considered the effect of a luminous transient destroying the dust, the wind density also cannot have significantly ($\gtorder 20\%$) dropped between the epoch of the HST observations (2002 November) and the last LBT observation (2022 February) since this would produce similar changes in the optical depth. This limit on the peak luminosity of any transient is quite conservative because it assumes that the obscured source is just an unperturbed progenitor.  In reality, the star will be more luminous and hotter for an extended period of time (e.g., \citealt{Fuller2017}, \citealt{Tsuna2023}), so this is the maximum amount of dust destruction needed to have produced a visible signal in the LBT data.   Unfortunately, the last usable LBT observation was from 2022 February, and so we can say nothing about possible transients in its last $\sim$1.3~yr of life.  Nevertheless, these data demonstrate how surveys like the search for failed SNe with the LBT can also be used to study of pre-SN variability and SNe in general. 

\section{Acknowledgements}

J.M.M.N. and C.S.K. are supported by NSF grants AST-1814440 and AST-1908570.

This work is based on observations made with the Large Binocular Telescope.  The LBT is an international collaboration among institutions in the United States, Italy and Germany. LBT Corporation partners are: The University of Arizona on behalf of the Arizona Board of Regents; Istituto Nazionale di Astrofisica, Italy; LBT Beteiligungsgesellschaft, Germany, representing the Max-Planck Society, The Leibniz Institute for Astrophysics Potsdam, and Heidelberg University; The Ohio State University, representing OSU, University of Notre Dame, University of Minnesota and University of Virginia. 

\section*{Data availability} 

The pre-SN LBT lightcurve of SN\,2023ixf shown in Figure~\ref{fig:lc} is available upon request. All other data is publically available.


\bibliographystyle{mnras}
\bibliography{bibliography}

\label{lastpage}
\bsp	
\end{document}